\documentclass{article}

\usepackage[english]{babel}
\usepackage{authblk}
\usepackage[letterpaper,top=2cm,bottom=2cm,left=3cm,right=3cm,marginparwidth=1.75cm]{geometry}

\usepackage{amsmath}
\usepackage{graphicx}
\usepackage[colorlinks=true, allcolors=blue]{hyperref}

\title{Decoding the News Media Habits of Disinformation Spreaders}
\author[1, 2]{Anna Bertani}
\author[1]{Valeria Mazzeo}
\author[1]{Riccardo Gallotti}

\affil[1]{Fondazione Bruno Kessler, Via Sommarive, 18, 38123, Povo TN}
\affil[2]{University of Trento, Department of Information Engineering and Computer Science, Via Sommarive 9, 38123, Povo TN}

\begin{document}
\maketitle

\begin{abstract}
In the digital era, information consumption is predominantly channeled through online news media disseminated on social media platforms. Understanding the complex dynamics of the news media environment and user’s habits within the digital ecosystem is a challenging task that requires at the same time large bases of data and accurate methodological approaches. This study contributes to this expanding research landscape by employing network science methodologies and entropic measures to analyze the behavioral patterns of social media users sharing news pieces and dig into the diverse news consumption habits within different online social media user groups.
Our analyses reveal that users are more inclined to share news classified as fake when they have previously posted conspiracy or junk science content, and vice versa, creating a series of "misinformation hot streaks". 
To better understand these dynamics, we used three different measures of entropy to gain insights into the news media habits of each user, finding that the patterns of news consumption significantly differ among users when focusing on disinformation spreaders, as opposed to accounts sharing reliable or low-risk content. Thanks to these entropic measures, we quantify the variety and the regularity of the news media diet, finding that those disseminating unreliable content exhibit a more varied and at the same time a more regular choice of web domains. 
This quantitative insight into the nuances of news consumption behaviors exhibited by disinformation spreaders holds the potential to significantly inform the strategic formulation of more robust and adaptive social media moderation policies.
\end{abstract}

\section{Introduction}

In the contemporary digital landscape, the spread of misleading and false information is becoming one of the most challenging issues that our society has ever faced before \cite{lazer2018science, zarocostas2020fight}. With the advent of online social media \cite{osatuyi2013information}, the access to information has drastically increased, allowing ideas and news to circulate rapidly and to reach a global audience. This shift has prompted individuals to actively seek and consume information directly through online social media platforms. In contrast to the pre-digital era, where people relied on traditional news media outlets and printed journals for information, the digital age has multiplied and simplified access to these sources. News media journals, once accessible primarily through physical copies, are now readily available with just a few clicks, marking a fundamental change in the way individuals obtain and engage with information \cite{westlund2015accessing}. In this context, online social media serves as a digital space where people can look for online news article as well as come across articles posted by other users \cite{nielsen2014relative}. New technologies have increased the possibilities for how people receive and send information \cite{westerman2014social}, 
disrupting completely the reading habits of online users \cite{mazzeo2022investigating}. 
The increasing number of accessible online news media articles has surely contributed to the problem of information overload \cite{renjith2017effect, rodriguez2014quantifying, koroleva2010stop, roetzel2019information}, also caused by the absence of a proper content regulation on the Internet \cite{rodrigues2020covid19}, that might affect the different spheres of democracy, such as freedom of thoughts, belief, expression \cite{mazzeo2022investigating}. 
This situation might be exacerbated during highly debated and contentious topics, such as the political elections \cite{grinberg2019fake} or the Covid-19 Pandemic \cite{shahi2021exploratory}, which have been characterized by a notably increase of dissemination of unreliable content \cite{caldarelli2021flow}.
Based on these premises, the aim of this work is to investigate the intricate news media digital landscape by analyzing the type of news media sources that dominate the online conversation. In addition, our interest focuses on the characterization of the behavioural patterns that each user displays in the selection of the news media types and web-domain chosen to read and share online. While extensive research has been conducted on news media consumption \cite{dutta2004complementarity}, particularly concerning the transformative impact of the Internet on information acquisition through websites and online articles, a notable gap exists in quantifying the diversity of users' news media diets within the digital and social landscape. Consequently, our interest goes beyond a mere descriptive analysis of user behavior in accessing and staying informed through online social media. We aim to introduce two measures able of capturing the \emph{variety} of news media consumption and the \emph{regularity} in accessing this information, drawing inspiration by two distinct entropic measures.
A deeper insight in users' news media habits might shed light on the extent to which there is a variety in choosing different news media journal or if they are more likely to stick to a few of them.  Decoding which sources online users rely on and share might raise our knowledge about how misinformative and disinformative content spread online and how users are more likely to be attracted by certain web-domains rather than others over time.

\section{Materials and Methods}
\subsection{Overview of the dataset}
We analyzed online social media data gathered by the Covid-19 Infodemic Platform \cite{covid19infodemics}, a comprehensive database containing messages posted on Twitter throughout the Covid-19 Pandemic. Specifically, we considered about 9,157,655 messages containing at least one URL (of which we were able to classify 2,549,226) posted in 2020.

We only consider the original messages, defined as tweets, thus not including in our analysis replies, retweets or quotes. To ensure a robust sample, we considered in our analysis only 25 countries with an average of at least 500 tweets per day collected in our platform. The countries included in our analysis are those associated with the following iso3 codes: COL, IRL, ITA, VEN, TUR, SWE, ESP, BRA, NLD, DEU, JPN, NGA, POL, CHE, AUS, AUT, CAN, CUB, ECU, GRC, MEX, PRT, ROU, SLV, ZAF. In total, we gathered information about messages posted by 211,493 users from the beginning of February until the end of 2020.

\subsection{News Reliability}
In our analysis, we evaluate the reliability of news domain by matching the URLs included in the textual content of messages against including information manually verified from various publicly accessible databases that cover scientific and journalistic sources. We specifically utilized data from MediaBiasFactCheck \cite{mediabiasfact}, an organization maintaining an extensive and regularly updated database. Their methodology involves a comprehensive evaluation of the ideological leanings and factual accuracy of media and information outlets, employing both quantitative metrics and qualitative assessments. The classification proposed by MediaBiasFactCheck has been further integrated with a number other publicly available sources (see \cite{gallotti2020assessing} for more details), identifying a total of 4,417 domains. In the process of developing  this domain classification, we also assessed the language representation of the classified web domains to take into consideration the intrisinc multilingual and multicultural aspects of our analysis. As demonstrated by Gallotti et al. \cite{gallotti2020assessing} comparing  web traffic statistics for the different countries and focusing on the top 50 most visited websites, the domains we classified match the top-tier websites across several countries of different native language, suggesting the reliability of the results for a comprehensive multilingual and multicultural analysis.

The URLs within the tweets were automatically scrutinized using this domain list, and each URL was categorized based on its source type, such as political, satire, mainstream media, science, conspiracy/junk science, clickbait, fake/hoax. Naturally, a fraction of domains was not present in our list: the most frequent cases being (i) shortened URLs for which the original link was not possible to reconstruct) and {ii) all the web-domains that have not been classified by external experts.

Building on the approach outlined in \cite{gallotti2020assessing}, we classified news sources as reliable (belonging to the Science or Mainstream Media categories), low-risk (Satire, Clickbait, Political, Other), or unreliable (Fake or Hoax, Conspiracy or Junk science). The distinction between web domains, which represent the actual URLs pointing to specific news media sources in the messages, and the categorized types within each domain is pivotal in our analysis, that wants to study at the same time the macro- and a microscopical perspective of the phenomenon of news "consumption" considering both the aggregation in news categories and the individual domains.

\begin{table}[h!]
    \centering
    \begin{tabular}{|c|c|}
        \hline
        News Category &  Level \\
        \hline
        Mainstream Media (MSM) & Reliable \\
        Science & Reliable \\
        Satire &  Low-Risk  \\
        Political & Low-Risk \\
        Clickbait & Low-Risk \\
        Other & Low-Risk \\
        Fake/Hoax & High-Risk \\
        Conspiracy/Junk science & High-Risk\\
        \hline
    \end{tabular}
    \caption{Classification of the news media categories based on their level of reliability.}
    \label{tab:simple}
\end{table}

\subsection{Building the information networks}

Each user is characterized by a sequence \emph{X} of web-domains and news media categories shared during the time considered (See Figure \ref{fig:infografica}). Based on these sequences, we reconstructed the undirected weighted networks where the nodes are web-domains and news media types and the weight assigned to the edge is the frequency of co-occurrence, aggregated by the total number of users. The network is built based on the subsequent posting of web-domains and news types to gain insights about the probability of choosing among different news media categories and web-domains. 
Moreover, we quantified the number of times users iteratively share the same web-domain and news media type, inspecting in depth the role of \emph{self-loops} in the network to gain a better understanding of their appeals towards certain web-domain or news media categories.

\subsection{User categories}

Given a sequence of different web-domains and news media categories (See again Figure \ref{fig:infografica}),  we studied the news media diet of each user based on the variety of web-domains and news categories shared.
It is of our interest to explore whether there are any differences in the selection of specific domains and news media categories based on the type of users considered. For this reason, we distinguish users into 5 categories: reliable-low risk (not sharing any news coming from fake or conspiracy news sources) and high risk users (sharing at least one fake or conspiracy news), classified further based on their increasing engagement with unreliable sources (more than 1, 5, 10 or 100 tweets in total).

\subsection{Entropy}

Entropy is a key measure to quantify the uncertainty of sources of information and it has played a central role in the field of communication theory \cite{shannon1948mathematical}. 
We employed three different measures of entropy to explore to what extent there is a diversity in the choice of web-domains and news media categories, considering different types of users.

First, the random entropy $S_{rand}=\log_{2}N_{i}$, where $N_{i}$ is the number of different characters in the sequence considered, in this case representing distinctive web-domains or news media types shared by user \emph{i}. This measure describe the size of the ensemble of different news shared by a user, and is useful to illustrate the predictability of the user's news media diet if each web-domain or news media category is shared with equal probability. 
Second, the temporal-uncorrelated entropy, computed as the Shannon entropy  \[
S_{unc} = -\sum_{j=1}^{N_i} p(x_j) \cdot \log_2(p(x_j))
\]
where $p(x_{i})$ is the historical probability that a web-domain $x_{i}$ was shared by user \emph{i}. This second measure thus allows us to characterize the heterogeneity of the news media diet.
Lastly, actual entropy, $S$,
depends not only on the frequency of web-domains chosen, but also on the order in which each web-domain is shared, capturing the temporal order present in a user's news media diet \cite{ziv1977universal}.
By definition \cite{song2010limits}, these three measures should respect the following inequality \[S \leq S_{unc} \leq S_{rand}\]
In on our data, this inequality did not hold in some cases for the actual entropy $S$. The Lempel-Ziv 
algorithm is an accurate estimator for the real entropy  on string sufficiently long\cite{wyner1989some, gao2003lempel} but this is not always the case in our data . For this reason, we decided to take into account only users who had at least three different web-domains and news media categories in order to exclude flawed entropy estimates from our analysis.

To complete our analysis, we also calculated the number of repetitions of web-domains and fact-type over the total number of potential couples over the length of each sequence in order to explore the weight of these repeated sequences of web-domains and news media categories shared by each user.

.
More specifically, given a sequence $X$ of length $L$, where $X= [x_1, x_2, \ldots, x_n]$, we counted the number of repeated pairs $x_j = x_{j+1}$ over all the total number of potential pairs $L-1$, to gain insights into the behavioral patterns in the sharing information, considering different type of users.
\section{Results}

We consider the sequence of messages containing an URL and the corresponding news media category for more than 211,493 users accounts, which have posted at least two different original tweets during 2020 (\ref{fig:infografica}). Each sequence has length \emph{L}, and has a specific number \emph{N} of unique domains posted. In this work, we decided to remove all users having posted only a unique web-domain during the period considered. These users accounts represents the 24\% of our sample, remaining with 160,228 unique users. In a second moment, as discussed above, we decided to remove also all users who shared only three different types of web-domains or news media categories in order to preserve the inequality of the three entropy measures \cite{song2010limits}, remaining with almost 18,000 unique users.

\begin{figure}[h!]
\includegraphics[width=14.5cm]{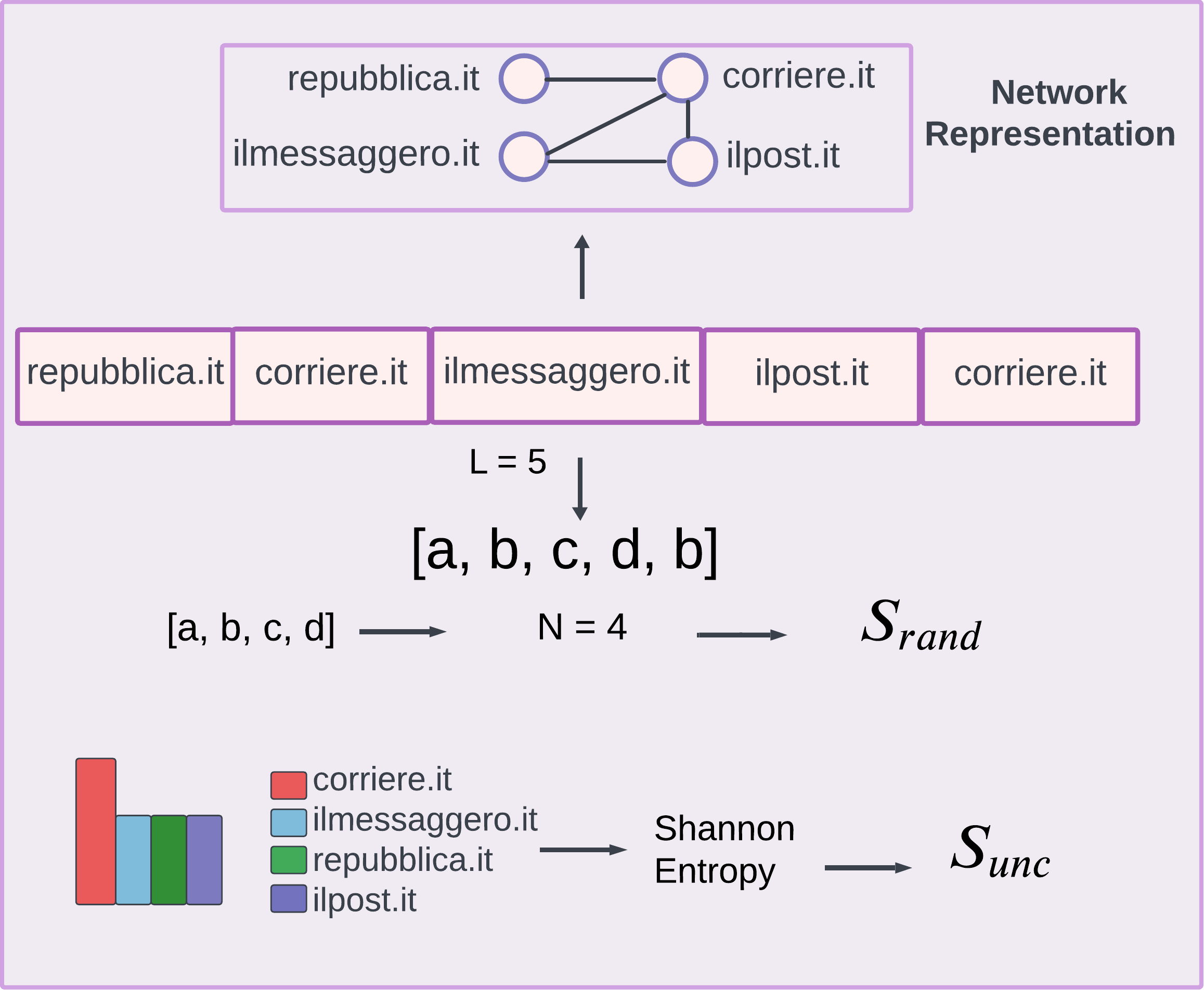}
\caption{{\bf The News Media Diet.} 
{Network and List of the sequence of web-domains posted by a random user on Twitter. We defined the length of each sequence with the term L. Each sequence is decoded in a list of identified letters. On these lists, we calculate the entropy value to get information about the variety of each sequence. In particular, we performed both random entropy and Shannon entropy calculations for each sequence.}\label{fig:infografica}}
\end{figure}

We first characterize the sequences. Figure \ref{fig:distribution} shows the probability distribution function of the sequence length $L$ of (Panel A) web-domains,
considering the entire dataset (Total), the dataset containing only those users who have at least 2 or more unique domains shared ($N>1$), and the different types of users we introduced in the method section.
Panel B shows the distribution of 
the number of tweets associated with 
each news media category in our dataset. The results confirm that mainstream media and "other" category are the main news media categories shared. Panel C reveals the distribution of
the number of tweets associated with 
the different news types shared by one particular user who had shown a great activity during the time considered.

This user was classified as \emph{Reliable/Low Risk} since he/she did not post any URL associated to conspiracy and fake news sources.

\begin{figure}[h!]
\includegraphics[width=14.5cm]{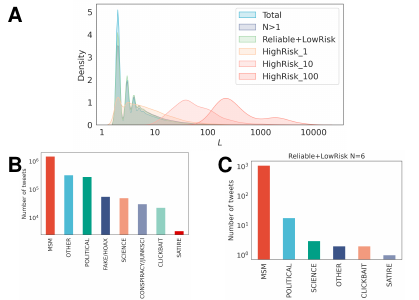}
\caption{{\bf Overview of the dataset. } 
{\bf (A)} Distribution of the length sequence of web-domains respectively for the entire dataset, for users having posted more than 2 distinctive web-domains (N>1), and users defined according the categories of reliability.
{\bf (B)} Distribution of the number messages classified in one of the eight categories of news media types.
{\bf (C)} Distribution of the news media categories used by a user particularly active, which has posted reliable and low-risk content, with 6 distinctive categories of news.}\label{fig:distribution}
\end{figure}

To give an overview of the news media environment as a whole, we reconstructed the undirected network of interactions among the web-domains shared by each user, where the nodes are represented by the web-domains and the weight is given by the number of times those couples of web domains have appeared in the sequence of posting. The network has been built at a coarser granularity having media categories as nodes in order to describe the inter-relations of web-domains of different categories posted. Figure The panel A (\ref{network}) shows that it is more likely to have posted a message containing a mainstream media source and news with a strong political bias, while panel B displays a heatmap quantifying the mean value of the relative value of the weight among the different news media categories. 

It is expected that links connecting the larger categories, MSM, political and Other, would have have larger weights and would appear dominantly in the network we built. To compensate the effect size of the web-domains distributed across the different news categories, we compare our findings with a null model that takes into account the relations normalized by the proportions of news that belong to each category, assuming a random sequence of news shared. Differently from the previous analysis, we found that, after compensating for the categories dimensions, it emerges a strong tendency among users to alternate between fake and conspiracy news, as shown by Panel C and by the corresponding values presented in Panel D. Differently, the relationship between the category science and the other ones is more equally distributed, as shown by the panel D. 

\begin{figure}[h!]
\includegraphics[width=14.5cm]{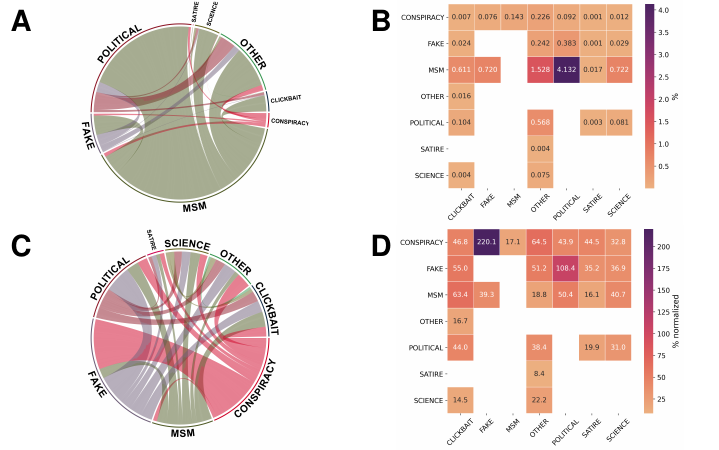}
\caption{{\bf The News Media Environment. } 
{\bf (A)} Weighted networks of the interactions among different news media types (also known as intra-relations) for 25 countries in 2020. 
{\bf (B)} Heatmap showing the corresponding value of the inter-relations among the eight categories.
{\bf (C)} Weighted networks normalized comparing with a null model accounting for the proportion of the number of news belonging to each category. The chances of sharing conspiracy and fake news by the same users is much higher than the strong relation observed between mainstream media and political news (Panel A).
{\bf (D)} Heatmap showing the corresponding normalized value of the inter-relations among the eight categories.\label{network}}
\end{figure}   

In light of this strong relation between fake and conspiracy news, we refer to the users who have posted fake or conspiracy news as \emph{high risk}, as already introduced in the methods section.

Moving from this first result, we decided to get a better insight of the intra-relations among domains and news media classes, also called \emph{self-loops}. The analysis displayed in Figure \ref{fig4}) reveals that the distribution of mainstream media is much broader with respect to the other classes when considering the web-domains chosen. Panel A highlights that users sharing MSM articles are more likely to be further sharing other articles from the same news category than users sharing other categories of media. Differently, the category ``Other'', encompassing general content that is not easily classified, such as videos on Youtube or posts on Instagram, shows that the percentage of domains chosen is limited to a few within this category. Conversely, Panel B shows that users sharing MSM have a lower tendency to repeatedly share of the same particular domain, while repetitions of the same domain are more frequent for sequences of Conspiracy theory or Clickbait content.

Beyond this analysis about the general patterns of news sharing in  online media, our focus extends to a more detailed exploration of users’ news consumption habits. In this regard, we classified each user into distinct groups: those not sharing fake and/or conspiracy news, and those showing different levels of unreliability (\emph{High Risk}, 1, 5, 10 and 100) based on their increasing engagement with high-risk news (see Methods).
In this context, we assigned to each user the three measures of entropy to better understand the news media diet and to gain insights into the potential differences between different kind of users.

\begin{figure}[h!]
\includegraphics[width=14.5 cm]{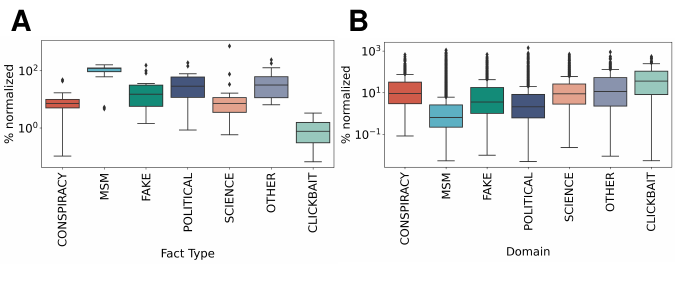}
\caption{{\bf Self-loops of web-domains and type of news shared. } 
{\bf (A)} The percentage distribution of the number of self-loops for different users, pointing to the same web-domains, respectively for all
the categories considered.
{\bf (B)} The percentage distribution of  the number of self-loops of different users, pointing to the same category of news, regardless the different web-domains shared.
\label{fig4}}
\end{figure}

\begin{figure}[h!]
\includegraphics[width=14.5cm]{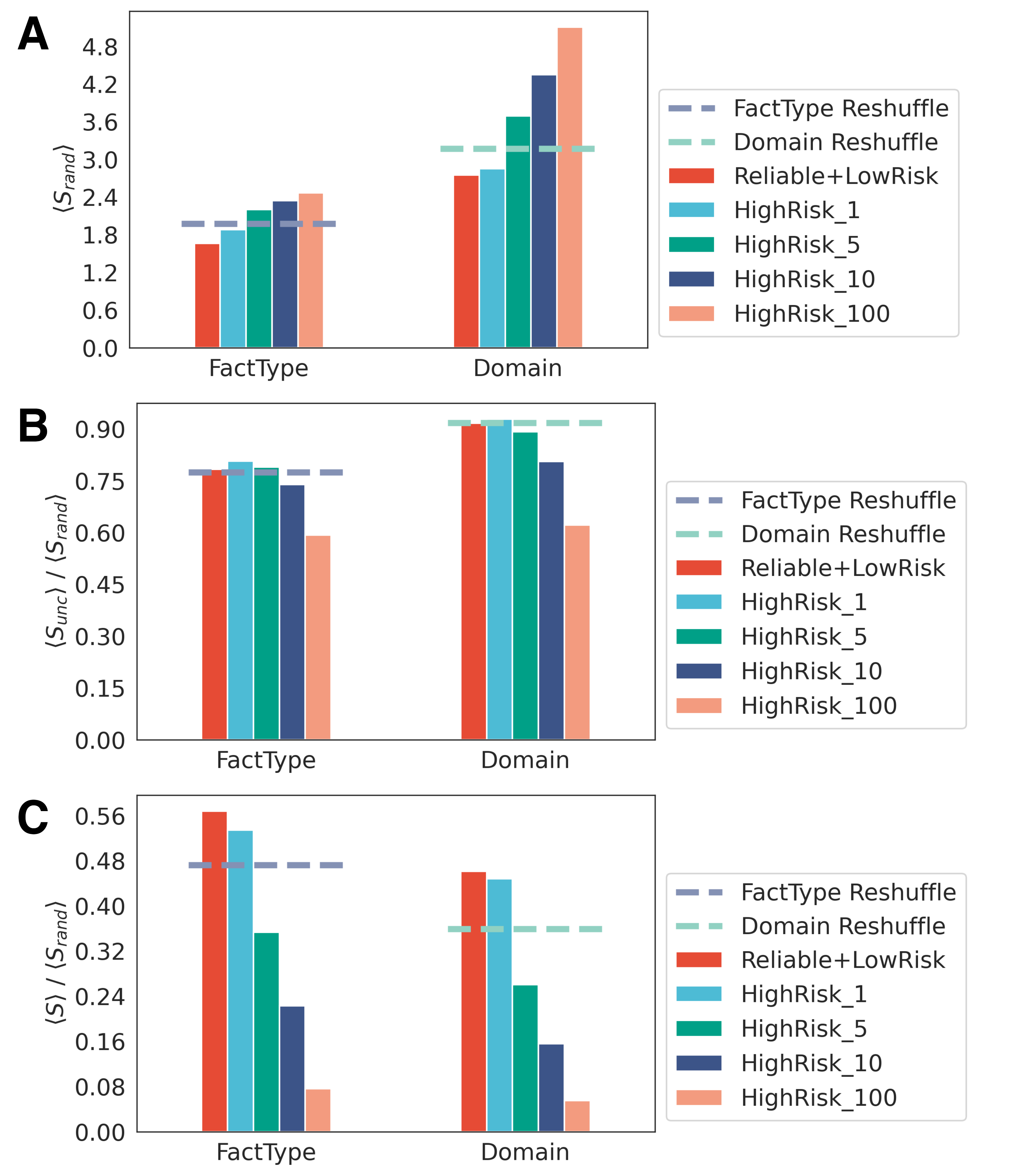}
\caption{{\bf News Media Diet for different type of users. } 
{\bf (A)} Random Entropy $S_{rand}$ calculated for different type of accounts: users posting reliable or low-risk content and users posting different levels of high risk (conspiracy/junk science and/or fake/hoax) content.
{\bf (B)} Shannon Entropy $S_{unc}$ calculated on the domain and the type of news shared by different type of users accounts: users posting reliable or low-risk content and users posting different levels of high risk (conspiracy/junk science and/or fake/hoax) content.
{\bf (C)} The actual entropy $S$ calculated on the domain and news media categories shared by different type of users: those posting reliable or low-risk content and those posting different levels of high risk (conspiracy/junk science and/or fake/hoax) content.
\label{entropy}}
\end{figure}  

Figure \ref{entropy} shows the distribution of (Panel A) the random entropy $S_{rand}$, (Panel B) the compression rate between the uncorrelated entropy $S_{unc}$ and the random one $S_{rand}$, and (Panel C) the compression rate between \emph{S} and $S_{rand}$ for different type of users. To validate our results, we calculated the proportion of web domains and news media types classified as High Risk relative to the total within their respective classes. This allowed us to extract a control sample of random web domains with a similar size to that of High Risk domains. 
We also tested that our results are not influenced by any potential selection biases associated with the definition of our user categories by randomly reshuffling the news categories in our tweets ensemble and re-assigning users in this shuffled categories. This process resulted in a constant value of entropy across all categories confirms absence of bias. These mean entropy values obtained from reshuffled samples (indicated by the dashed lines) are consistently lower than those of the four high-risk groups, especially for the $S_{rand}$, which we interpret as a measure of variety of the news media diet. Users sharing reliable or low-risk content seems to have a more restricted news media diet not only when considering the news media categories chosen but also when looking at the number of domains shared. Differently, users  classified as high risk shows higher values of random entropy as their engagement with these types of content increases, indicating that the consumption and sharing of high risk content can be associated with wider exploration of a larger number of different news categories and domains shared. 

The compression rate between $S_{unc}$ and $S_{rand}$ appears relatively flat for all the types of users considered, but here the pattern for high-risk users being more compressible than the random expectations. This tendency is even stronger in the third panel (C), which displays the compression rate between $S$ and $S_{rand}$ as a measure of regularity of the news media diet. In this case, the compression rate for high-risks user is much lower than the second panel (B). This indicates that the same heavy consumers and spreaders of high risk content that we see have a tendency of exploring a larger number of news categories and domains, displaying at the same time a stronger regularity in their choice of web-domains and news media categories.

In order to get deeper insights into this counter-intuitive finding, we study the relations between $S_{rand}$ and the percentage of news media types and web-domains repeated, shedding more lights on users' tendencies to either explore new web domains and categories or stick to a few of them. Figure \ref{perc_domain_fact} illustrates the values of $N$ of different news media category (or web domains) shared against the percentage of news media repeated for reliable/low risk and high-risk users, as defined in the methods section . 
The Panel (A) demonstrates that as the number of news media type increases, the percentage of fact-type repeated grows. The last distribution characterizing the users with the highest value of $S_{rand}$
entropy are the ones with a higher number of mainstream media and political journals repeated. The expectation of repeating the same fact-type assuming a homogeneous distribution is instead expected to be decreasing for growing $N$, as it is trivially given by the ratio $1/N$, highlighted by the dashed line in all panels. The distribution of repeated news media categories for users posting six unique news media categories is therefore displaying a different tendency and is significantly higher than the random expectation for larger $N$. 
However, the same behavior is not shown when looking at users posting high-risk content, as shown in the Panel B. Indeed, the distribution of the percentage of news media types shared is slightly higher for users with a lower $S_{rand}$. The last distribution is represented by people having posted all the possible news categories (\emph{N = 8}) 
Besides the news media classes, we analyzed the percentage of web-domains repeated with respect to the number of unique domains shared, respectively for reliable or low risk users (Panel C) and high risk users (Panel D). In this case, the distribution of domains repeated shows to be fixed at the 25 \%, independently from the size of  \(\langle N \rangle\) both for reliable and low-risk and high risk users. Interestingly, this result indicates that the percentage of repeated domains is the same regardless of the type of users and the number of domains or news media categories shared.

\begin{figure}[h!]
\includegraphics[width=14.5cm]{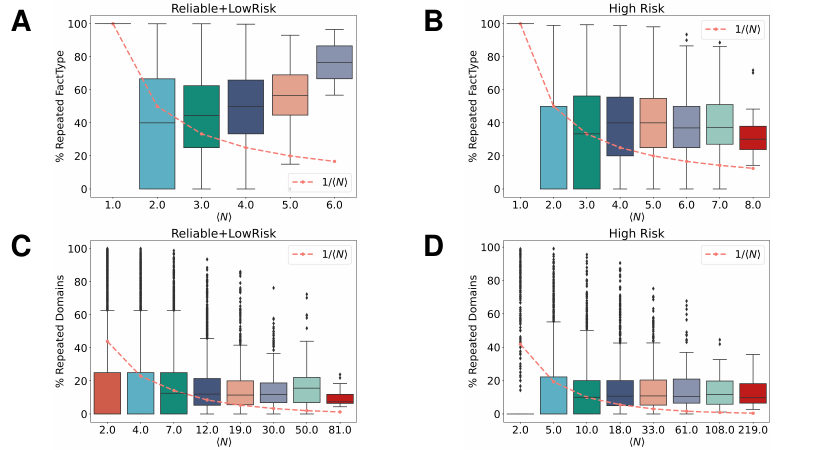}
\caption{{\bf Fraction of repeated news media categories and web-domains for users with different $N$.} In these boxplots we represent the percentage of news media categories (Panels A,B) or web domains (Panels C,D) repeated in subsequent tweets with respect to the total length of the sequence associated to an individual user. Each boxplot is aggregated over users with the same $N$. The dashed line shows the random expectation of repeated news media categories with respect to the increasing number of $N$ unique news media category. 
{\bf (A)} Patterns of news media categories, aggregated for user classified as having posted only reliable and/or low-risk content on Twitter. The box-plots here are eight since we have 6 distinctive categories as two are filtered out.
{\bf (B)} Patterns of news media categories, aggregated for users having posted also high-risk content. The box-plots here are eight since we have all 8 distinctive categories of news media types. 
{\bf (C)} Patterns of web-domains, aggregated for by users classified as having posted reliable and/or low-risk content. 
{\bf (D)} Patterns of web-domains, aggregated for by users classified as having posted also high-risk content in their sequence of messages posted. 
\label{perc_domain_fact}}
\end{figure}




\section{Discussion}

Our research reveals that the news media digital landscape is a complex environment where users tend to be attracted and thus to repeat specific web-domains or news media categories. In particular, when we inspect at the news media environment, we found that there is a strong relation between the probability of sharing mainstream media journals and political media. This relation is just apparent, because if we compare our results with a null model that takes into account the effect size of each category, our analysis highlights how users sharing fake or hoax news are more likely to alternate with conspiracy and junk science type of news, illustrating the tendency for unfortunate misinformation hot streaks. This result also suggests how much it is easy to fall down in this vicious circle where alternating fake and conspiracy news becomes the norm. Differently, the news categories representing the most reliable ones (Mainstream Media and Science) do not show this strong relationship, leaving the most unreliable categories at the center of the news media digital landscape. This tendency of alternating fake and conspiracy news has raised up the need of better understanding which are the characteristic features of each user, specifically when we consider different type of users. In this scenario, one might suggests that different type of users based on their interactions with different type of content have a completely different news media habits. This is what our research shows. In this context, we introduced two different measures to characterize and quantify the users' behaviour in the way they access and retain information online, shaping their news media diet. The \emph{variety} of the so-called news media diet is given by the $S_{rand}$, while the \emph{regularity} is calculated by the compression rate between  $S$ and $S_{rand}$. By employing these measures of entropy, we found that reliable or low-risk users have a more restricted news media habits, when searching for news media categories and web-domains. On the contrary, high-risk users at different levels reveal to have a more varied news media diet, especially for users sharing at least 100 type of high-risk web-domains. 
Interestingly, the ratio between the $S_{unc}$ and the $S_{rand}$ demonstrates that reliable or low-risk users have less compressible sequences if compared with high-risk users. This is even more explicit when we observe the compression rate between the real entropy $S$ and the random entropy $S_{rand}$, which highlights how high risk users display a remarkable regularity in their news consumption, even if they are characterized by having a more diverse diet if we just consider the random entropy which takes into account only the number of unique web-domains and news media categories. 
Getting deeper into this study, we found that the repetitions of news media categories is in general higher than expected if we look at the users characterized by a more huge variety of news media categories selected. Interestingly, this effect is stronger for reliable and low-risk users. In that context, we observed that those users are usually repeating also more often political news media categories and represent particularly active users. On the contrary, the repetitions of web-domains seems independent by the variety of the news diet, whether the user is a reliable or low-risk or  high-risk one, showing that there is the same probability of sharing web-domains, independently by how much the user has a wider news media habits. 
This work has shed lights on the news media behavior and patterns of the users during highly contentious events, such as the Covid-19 Pandemic, revealing how much certain type of users have more varied but at the same time regular news diet as they are more likely to share and to repeat the same news media sources. In particular, we found that this happens for users heavily sharing high-risk disinformation sources. 
At the same time, we also observed from an aggregated perspective how news media digital environment is shaped according the strong relationship of fake and conspiracy news: highly debated and unprecedented topics can generate debates driven by fake and conspiracy news which can monopolize the attention of particular users. These users with a greater tendency to engage with these high-risk content are also those more likely to have a wider content base but also the ones more likely to express a strong regularity in the way they share news media content online over-time. 
These findings underscore the imperative for a more nuanced comprehension and proactive analysis of the ramifications on online social systems, alongside a discerning evaluation of the offline risks engendered by the dissemination of streaks of misinformative content within the digital sphere. This is certainly a topic that merits additional analyses and holds significant potential for enhancing our comprehension of the news media habits of users during contentious debates. 





\vspace{6pt} 




}
\section*{Acknowledgements}
R.G. acknowledges the financial support received from the European Union's Horizon Europe research and innovation program under grant agreement No 101070190.
A.B and R.G. acknowledge the support of the PNRR ICSC National Research Centre for High Performance Computing, Big Data and Quantum Computing (CN00000013), under the NRRP MUR program funded by the NextGenerationEU.

\bibliographystyle{plain}
\bibliography{sample}

\end{document}